\documentclass[12pt,a4paper,DIV12]{scrartcl}
\usepackage[utf8x]{inputenc}
\usepackage[T1]{fontenc}
\usepackage{lmodern}
\usepackage[british]{babel}
\usepackage[pdftex]{graphicx}
\usepackage{hyperref}
\usepackage{color}
\usepackage{textcomp}
\usepackage{subfigure}
\usepackage{amsmath}
\usepackage{amssymb}
\usepackage{amsfonts}
\usepackage{lscape}
\newcommand{\arxiv}[1]{arXiv:\,\href{http://arxiv.org/abs/#1}{{\tt #1}}}

\title{
  {\vspace{-20mm}\normalsize
   \hfill\parbox[b][30mm][t]{35mm}{\textmd{MS-TP-14-38}\\
                                   \textmd{DESY 14-210}}}\\[-18mm]
Influence of topology on the scale setting
\vspace*{2mm}
}

\author{G.~Bergner\\
\textit{\large Universit\"at Bern, Institut für Theoretische Physik}\\
\textit{\large Sidlerstr.~5, CH-3012 Bern, Switzerland}\\
\textit{\large E-mail: bergner@itp.unibe.ch}\\[3mm]
I.~Montvay\\
\textit{\large Deutsches Elektronen-Synchrotron DESY}\\
\textit{\large Notkestr. 85, D-22603 Hamburg, Germany}\\
\textit{\large E-mail: montvay@mail.desy.de}\\[3mm]
P.~Giudice, G.~M\"unster, S.~Piemonte\\
\textit{\large Universit\"at M\"unster, Institut f\"ur Theoretische Physik}\\
\textit{\large Wilhelm-Klemm-Str.~9, D-48149 M\"unster, Germany}\\
\textit{\large E-mail: munsteg, p.giudice, spiemonte@uni-muenster.de}
\vspace*{2mm}
}

\date{November 25, 2014}

\begin{document}

\maketitle

\begin{abstract}
Recently a new method to set the scale in lattice gauge theories, based on
the gradient flow generated by the Wilson action, has been proposed, and the
systematic errors of the new scales $t_0$ and $w_0$ have been investigated
by various groups. 
The Wilson flow provides also an interesting alternative smoothing procedure 
in particular useful for the measurement of the topological charge as a 
pure gluonic observable.
We show the viability of this method for
$\mathcal{N} = 1$ supersymmetric Yang-Mills theory by analysing the
configurations produced by the DESY-Muenster collaboration. For increasing flow
time the topological charge quickly approaches near-integer values. The
topological susceptibility has been measured for different fermion masses
and its value is observed to approach zero in the chiral limit. Finally, the
relation between the scale defined by the Wilson flow and the topological
charge has been investigated, demonstrating a correlation between these two
quantities.
\end{abstract}
\vspace{5mm}
\newpage

\section{Introduction}
Lattice regularisation allows nonpertubative investigations of quantum field
theories. The continuum space-time is discretised to a hypercubic finite
lattice with points separated by a distance $a$. The integral over all
possible configurations then has a mathematically well-defined meaning, and
Monte Carlo methods can be applied to approximate expectation values of
observables. The lattice spacing $a$ is an important dimensionful parameter
in the regularised theory; knowledge of its value is crucial to extrapolate
physical quantities to the continuum limit $a\rightarrow 0$, to test the
agreement with experimental data or simply to compare results between
different lattice actions. The value of the lattice spacing is implicitly
defined once a dimensionful observable, for instance the mass of a particle
in physical units, is chosen as input parameter to set the scale and to
match simulations done with different bare parameters.

It is important to determine the scale as precise as possible since its
error contributes a significant part to both statistical and systematic
errors of the lattice results, that propagates to the final physical predictions
of Monte Carlo simulations. Therefore, the observable used to set the scale
has to be chosen with special care. Various examples for this purpose have
been investigated during the last two decades. In particular, three of them
have been successfully tested for many different theories: the Sommer
parameter $r_0$ and the Wilson flow scales $t_0$ and $w_0$.

The Sommer parameter $r_0$ was first proposed in Ref.~\cite{Sommer:1993ce}
as the distance $r$ where the strong force between a static quark-antiquark
pair multiplied by the squared distance, $r^2 F(r)$, reaches some specified
value, typically $1.0$ or $1.65$. The Sommer parameter is a pure gluonic
observable in the sense that it requires only the computation of expectation 
values of Wilson loops. While this measurement is computationally inexpensive,
noisy signals affect the result for the interquark force at large distances 
where however lattice artefacts are small. 
Systematic errors arise when different choices of
smoothing procedures are used to improve the signal of $F(r)$
and when the fitting procedure is employed to extract the value of 
the parameters,
increasing the complexity of the measurement of the Sommer parameter $r_0$.

Recently a new method to set the scale by the parameter $t_0$ has been
proposed in Ref.~\cite{Luscher:2010iy}, based on the gradient flow 
generated by the Wilson or the Symanzik gauge field action
\cite{Narayanan:2006rf}. A closely related method based on the parameter
$w_0$ has been developed in Ref.~\cite{Borsanyi:2012zs}. In this paper we
compute the scale parameters $t_0$ and $w_0$ for the $\mathcal{N}=1$ SU(2)
supersymmetric Yang-Mills (SYM) theory and discuss
their systematic errors. Our calculations employ the configurations
generated by the DESY-M\"unster collaboration
\cite{Bergner:2012rv,Bergner:2013nwa,Demmouche:2010sf}. We show that, for
large flow times, correlations appear between the topological charge
and the scale setting quantities. As a consequence, unexpected finite 
volume effects can arise in the
computation of $w_0$ and $t_0$. We further show that a fine tuning of the
scales $t_0$ and $w_0$ can drastically reduce this effect without
introducing any further systematic errors. Similar analyses of the influence
of topology on the scale setting have been presented in
Ref.~\cite{Bruckmann:2009cv,Aoki:2008tq} for the Sommer parameter $r_0$.
Yang-Mills theories at fixed topology have been intensively studied in the
literature, see for example \cite{Brower:2003yx,Aoki:2007ka}.

\section{The Wilson flow}

%
The Wilson flow can be considered as a continuous generalisation of stout
smearing\nolinebreak\ \cite{Morningstar:2003gk}. The starting point is to introduce an
additional fictitious time $t$ as fifth dimension, in the course of which
the gauge fields $U_\mu(x)$ generated by Monte Carlo simulations are
``continuously smoothed''. The continuous smoothing procedure is specified
by the partial differential equation
\begin{equation}
 \frac{\partial}{\partial t} V_\mu(x,t) 
= - g^2 S_{\textrm{gauge}}(V_\mu(x,\tau))\, V_\mu(x,t)\,,
\end{equation}
similar to a diffusion equation, with boundary conditions
\begin{equation}
 V_\mu(x,t)|_{t = 0} = U_\mu(x)\,.
\end{equation}
Here $V_\mu(x,t)$ denotes the link variables at fictitious time $t$ and
$U_\mu(x)$ the original link variables.

The Wilson flow removes ultraviolet divergences and therefore local gauge
invariant operators defined at positive flow time are automatically
renormalised. Quantities constructed from the link variables $V_\mu(x,\tau)$
have a well-defined continuum limit and can be used to set the scale in
lattice simulations.

The scale $t_0$ has been introduced in Ref.~\cite{Luscher:2010iy} as the
flow time $t$ fulfilling
\begin{equation}\label{definition_t0}
 t^2 \langle E(t) \rangle = 0.3\,.
\end{equation}
Here the gauge energy $E(t)$ is defined as
\begin{equation}
 E=\frac{1}{4}G_{\mu\nu}^a G_{\mu\nu}^a\,,
\end{equation}
where $G_{\mu\nu}$ is a lattice version of the field strength tensor
$F_{\mu\nu}$ which, as usual, is specified by the antisymmetric clover plaquette.
The scale $t_0$ has the same dimension of
the inverse string tension, i.e.\ length squared.

The closely related scale $w_0$ has been introduced in
Ref.~\cite{Borsanyi:2012zs} as the square root of the flow time $t$ where
the condition
\begin{equation}\label{definition_w0}
 t \frac{d}{dt} t^2 \langle E(t) \rangle = 0.3
\end{equation}
is satisfied. $w_0$ has the dimension of a length, i.e.\ the same dimension
of the lattice spacing. It has been demonstrated that $w_0$ is less sensitive
to lattice artefacts than $\sqrt{t_0}$. According to
Ref.~\cite{Borsanyi:2012zs}, the difference between the application of the Symanzik
 and the Wilson gauge field action 
in the integration of the flow equation on the lattice is not relevant. 
In this work we apply the Wilson action since it requires a smaller computational effort. 
The Wilson flow has been numerically integrated using a Runge-Kutta scheme with steps
of length 0.01, as described in Appendix C of Ref.~\cite{Luscher:2010iy}.

\section{Measuring the scale setting quantities $w_0$ and $t_0$}

The gauge configurations have been generated by the {\em
Two-Step Polynomial Hybrid Monte Carlo (TSPHMC)} update algorithm
\cite{Montvay:2005tj,Scholz:2006hd} for the study of the hadron spectrum in
SYM with gauge group SU(2) \cite{Bergner:2012rv,Bergner:2013nwa,Demmouche:2010sf}. This theory
describes the interactions between gluons and their supersymmetric partners, 
the gluinos. The gluino is a Majorana fermion in the adjoint
representation of the gauge group. The Symanzik improved action has been
used for the gauge action, and the Wilson-Dirac operator with one-level
stout smeared links for the fermion action\footnote{The value of the stout smearing
parameter was $\rho=0.15$, see \cite{Bergner:2013nwa} for further details.}. We have
determined the Wilson flow scales for three different values of $\beta = 4/g^2$, where $g$ is the bare gauge coupling,
and many different values of the fermionic hopping parameter
$\kappa=1/(2m+8)$, where $m$ is the bare gluino mass.

The integration of the Wilson flow equation has been performed on every
sixth thermalised configurations\footnote{The individual configurations are
separated by 1 unit in HMC time, $T_{MC}=1$, the measured 
configurations are
separated by 6 units in HMC time. In the plots of the Monte Carlo history we will use
 therefore ``$T_{MC} \times 6$'' in the x-axis labels. The integrated autocorrelation time 
of the unsmeared plaquette is always below 1.5 in these units.}, and the results 
are summarised in Tab.~\ref{tab:w0results}. The scales $w_0$ and $\sqrt{t_0}$ show only a small
dependence on the gluino mass for a given $\beta$. Employing a mass
independent renormalisation scheme, the scales are extrapolated to the
chiral limit at zero renormalised gluino mass and the obtained value is
 used to set the scale at all gluino masses. In our calculations the
renormalised gluino mass is represented by the square of the (adjoint) pion
mass ($m_\pi$), which is defined in a partially quenched theory and can be measured 
with a reasonable precision. As shown in \cite{Munster:2014cja},
the gluino mass is proportional to the square of $m_\pi$.

%
\begin{figure}[ht]
\centering
\includegraphics[width=0.77\textwidth]{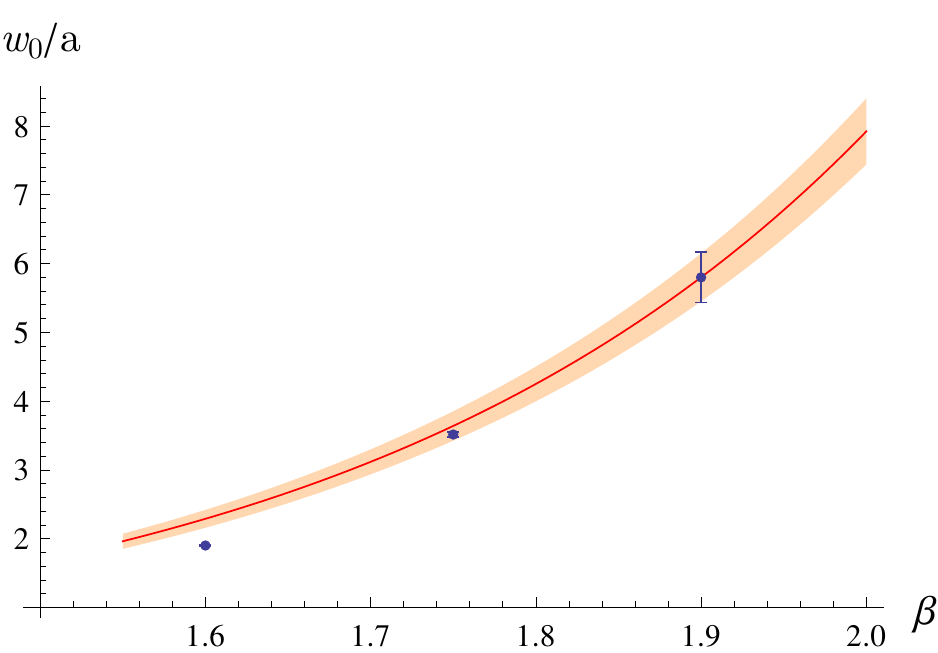}
\caption{Scaling of $w_0$ compared to the expected behaviour from the
$\beta$-function (red line). The orange band represents the statistical
error determined by Eq.~\ref{scalingw0}.}
\label{w0_scaling}
\end{figure}
\section{Matching the $\beta$-function}
The Callan-Symanzik-$\beta$-function has been determined for the
$\mathcal{N}=1$ SYM theory in Ref.~\cite{Novikov:1983uc} by instanton
calculations. The result
\begin{equation}
\beta(g) = \mu \frac{d}{d \mu}g(\mu) 
= - \frac{g^3}{16 \pi^2}\frac{3 N_c}{1-\frac{N_c g^2}{8 \pi^2}}
\end{equation}
is exact due to the non-renormalisation theorem~\cite{Novikov:1983uc}. 
The first two perturbative
coefficients are universal 
and scheme independent. 
The $\beta$-function can
be used to compare lattice results at different bare gauge couplings $g$. If
finite volume corrections and lattice discretisation errors can be
neglected, the Wilson flow parameters $t_0$ and $w_0$ are expected to scale
according to
\begin{equation}\label{scalingw0}
 \frac{w_0(g_1)}{w_0(g_2)} = \exp{(F(g_1) - F(g_2))},
\end{equation}
where the function $F(g)$ is the integral of the inverse of the
$\beta$-function:
\begin{equation}\label{scalingexp}
 F(g) = \int^g \frac{d g'}{\beta(g')}
= \frac{8 \pi^2}{3 N_c g^2} + \frac{2}{3} \ln g\,,
\end{equation}
up to an unessential integration constant.

For our case, $N_c = 2$, the scaling according to Eq.~\ref{scalingw0} has been
checked by taking the value of $w_0$ at $\beta=4/g^2=1.9$ as reference
point, see Fig.~\ref{w0_scaling}\,. The agreement with Eq.~\ref{scalingw0}
is rather good. 

The relative deviation from the scaling,
\begin{equation}\label{keq}
K = \frac{w_0(1.9)}{w_0(\beta)} 
\left(\frac{1.9}{\beta}\right)^{1/3}
\exp\left\{\frac{\pi^2 (\beta-1.9)}{3}\right\},
\end{equation}
is $K = 1.03(6)$ for $\beta=1.75$ and $K = 1.20(8)$ for $\beta=1.6$.
The larger deviation at $\beta=1.6$ is presumably due to lattice artefacts
and/or higher order terms in the lattice-$\beta$-function.

\begin{figure}[t]
\centering
\subfigure[$Q_\textrm{lat}(t)$]
{\includegraphics[width=0.48\textwidth]{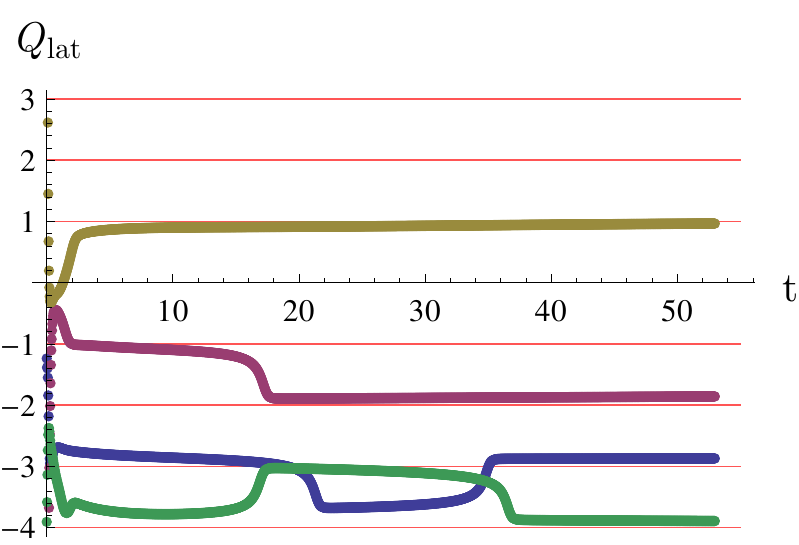}
\label{topological_charge_flow}}
\subfigure[PDF$(Q_\textrm{lat})$]
{\includegraphics[width=0.48\textwidth]{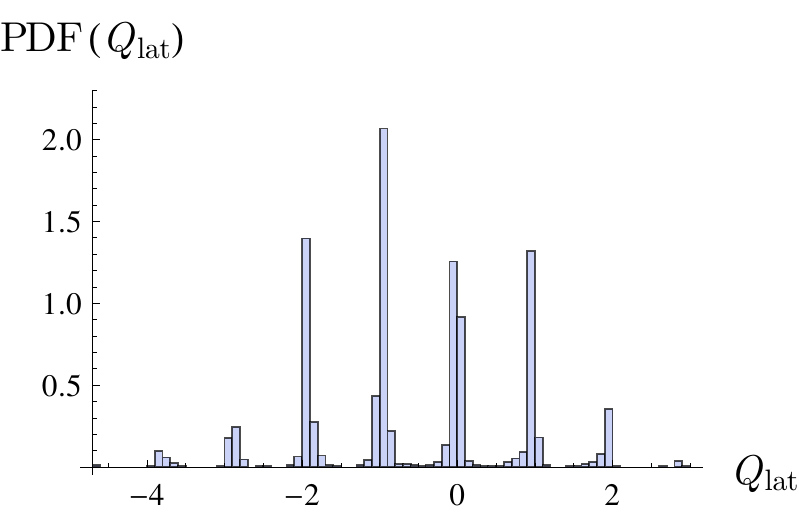}
\label{topological_charge_flow_dist}}
\caption{(a) Topological charge for four different configuration as a function of
the flow time $t$ on a $32^3 \times 64$ lattice, $\beta=1.9$ and
$\kappa=0.14435$. (b) Distribution of the topological charge at $t=50$
for the same run.}
\end{figure}
%
\section{Measuring the topological charge with the Wilson flow}\label{topological_charge_sec}

The topological charge is defined for a given field configuration in the 
continuum by the integral
\begin{equation}
 Q_\textrm{top} 
= \frac{1}{32 \pi^2} \int d^4 x\, \epsilon_{\mu\nu\rho\sigma} 
F^{\mu\nu}_a F^{\rho \sigma}_a.
\end{equation}
On the lattice we define the topological charge with the same antisymmetric
discretisation of the field strength tensor
as used for the flow equation:
\begin{equation}
 Q_\textrm{lat} 
= \frac{1}{32 \pi^2} \sum_x \epsilon_{\mu\nu\rho\sigma} 
G^{\mu\nu}_a G^{\rho \sigma}_a.
\end{equation}
This {\em lattice topological charge} is affected by ultraviolet
fluctuations, and its value is in general not an integer. A possible solution to
this problem is a smoothing procedure to suppress the short distance
fluctuations and to recover a well-defined topological charge in
the continuum limit \cite{Creutz:2010ec}. We have applied the
Wilson flow as smoothing procedure, as done in Ref.~\cite{Bonati:2014tqa}.

As shown in Fig.~\ref{topological_charge_flow}, for large enough flow time
$t$ the topological charge reaches a near integer value. Following
Ref.~\cite{Bonati:2014tqa}, we convert the raw lattice topological charge to
an integer using
\begin{equation}
Q_\textrm{top} = \textrm{round}{(\alpha Q_\textrm{lat}(t))},
\end{equation}
where the flow time $t$ is chosen to be
\begin{equation}
 t = \frac{1}{8}\left(\frac{L}{3}\right)^2
\end{equation}
where $L$ is the spatial extent of the lattice. This value of $t$ is chosen sufficiently large to remove the cut-off effects; but not too large to change the number of instantons and the final value of 
the topological charge~\cite{Smith:1998wt}\,.
The real constant
$\alpha$ is chosen to minimise the ex\-pec\-ta\-tion value
\begin{equation}
 R(\alpha) 
= \langle (\alpha Q_\textrm{lat} - \textrm{round}(\alpha Q_\textrm{lat}))^2 
\rangle.
\end{equation}
Near the continuum limit it is expected that $\alpha \approx 1$, i.e.\ the
distribution of $Q_\textrm{lat}$ is already centred near integer values
without requiring an additional multiplicative renormalisation, see
Fig.~\ref{topological_charge_flow_dist} (and Fig.~\ref{ralpha} in the Appendix).
\begin{figure}[t]
\centering
\subfigure[$\chi_{\textrm{top}}$ at $\beta=1.60$]
{\includegraphics[width=0.49\textwidth]{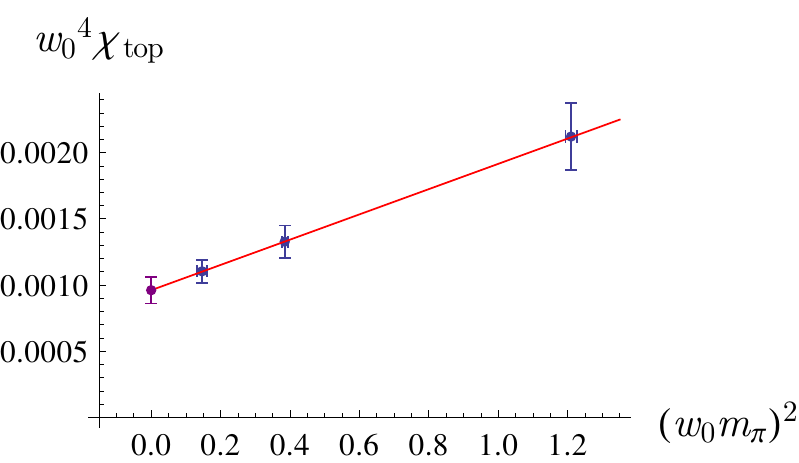}}
\subfigure[$\chi_{\textrm{top}}$ at $\beta=1.75$]
{\includegraphics[width=0.49\textwidth]{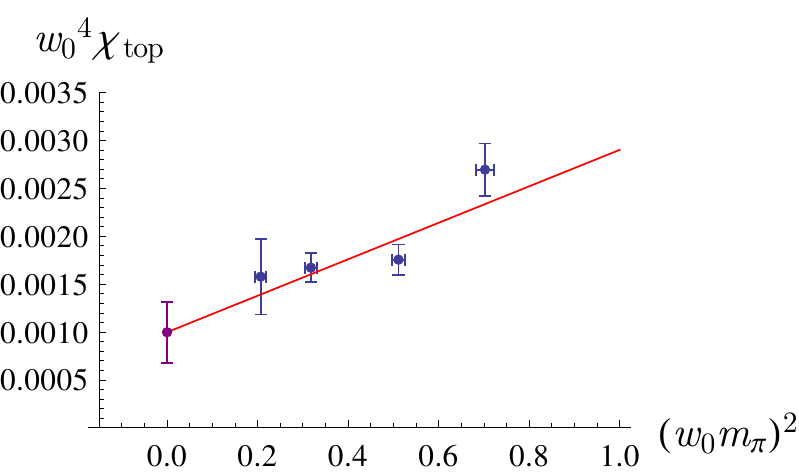}} 
\caption{Extrapolation of the topological susceptibility
$\chi_{\textrm{top}}$ to the chiral limit on a $32^3 \times 64$ lattice at (a) $\beta=1.60$ and (b) $\beta=1.75$.}\label{chitop}
\end{figure}
In addition, the topological susceptibility $\chi_{\textrm{top}}$, defined by
\begin{equation}
 \chi_{\textrm{top}} 
= \frac{\langle Q_{\textrm{top}}^2 \rangle 
- \langle Q_{\textrm{top}} \rangle^2}{V}\, = \frac{\langle Q_{\textrm{top}}^2 \rangle 
- \langle Q_{\textrm{top}} \rangle^2}{a^4 N_s^3 N_t},
\end{equation} 
where $V$ is the volume of the system, has been measured. The results
are shown in Tab.~\ref{tab:chiresults} and in Fig.~\ref{chitop}. The value of $\chi_{\textrm{top}}$
extrapolated to the chiral limit confirms the {\em topological suppression} for SYM mentioned in
Ref.~\cite{Bergner:2014iea}.

\section{Autocorrelation time of flow observables}

The autocorrelation time of the topological charge increases drastically
near the continuum limit and may even result in {\em topological freezing}.
This effect depends, however, on the chosen boundary conditions \cite{Luscher:2011kk,Luscher:2011qa}.
The scales $w_0$ and $t_0$ exhibit a very long autocorrelation time,
especially near the continuum limit, similarly to the topological charge.
Our configurations have been produced with the usual periodic boundary
conditions and we observe the expected increase of the autocorrelations.

The autocorrelation time should scale with the lattice spacing $a$
asymptotically as $a^{-z}$, where $z=1$ for Hybrid Monte Carlo (HMC)
algorithms. In our runs the lattice spacing is decreased roughly by a factor
$2.5$ between $\beta=1.6$ and $\beta=1.9$. The integrated autocorrelation time
$\tau(t_0)$ of $t_0$ at $\beta=1.9$ is, however, approximately twelve times
larger than at $\beta=1.6$, see Tab.~\ref{tab:w0results}.

Although the interval between $\beta=1.6$ and
$\beta=1.9$ is presumably not yet in the asymptotic regime, the variation of
$\tau(t_0)$ seems to indicate a value $z\gtrsim 2$ and a possible connection
of the topological charge with the flow observables used to set the scale.

\section{Correlations between topological charge and the scale $w_0$}

\begin{figure}[t]
\centering
\includegraphics[width=0.73\textwidth]{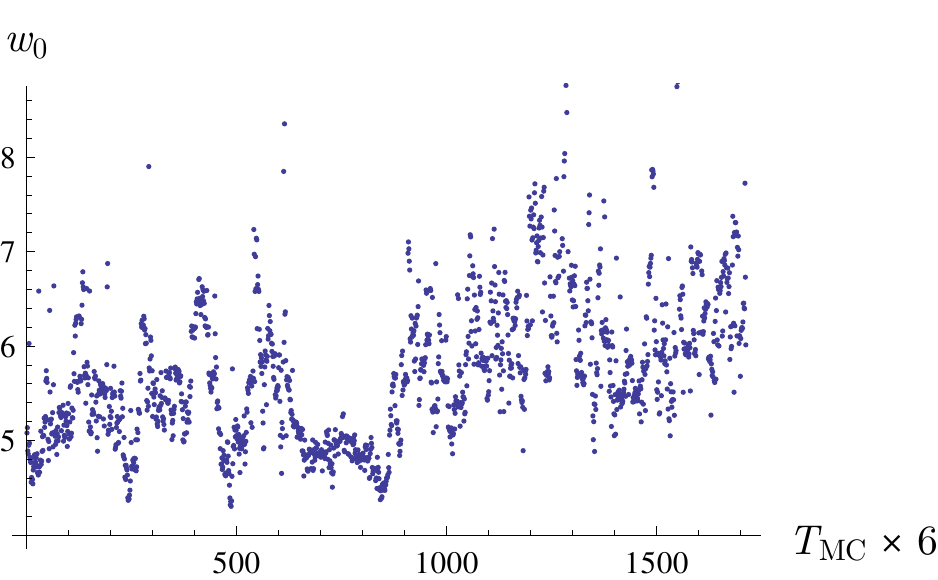}
\caption{Monte Carlo history of $w_0$ on a $32^3 \times 64$ lattice,
$\beta=1.9$ and $\kappa=0.14435$. $T_{MC}$ is the Monte Carlo time. $w_0$ is measured every sixth
configurations.}
\label{mc_history_w0_03}
\end{figure}
In order to investigate the nature of the long autocorrelation of $w_0$, we
have considered its Monte Carlo history. The scale $w_0$ can be defined for
a single configuration without the need of an ensemble average by
determining the flow time when the integrated flow matches the condition
specified by Eq.~\ref{definition_w0}\,. In Fig.~\ref{mc_history_w0_03} the
Monte Carlo history of $w_0$ is shown for a lattice $32^3 \times 64$ at
$\beta=1.9$ and $\kappa=0.14435$. One can see that the value of $w_0$ has
large fluctuations with a long period. In particular, very strong upward
spikes emerge.
\begin{figure}[ht]
\centering
\subfigure[$w_0^{0.1}$ versus $w_0^{0.3}$]
{\includegraphics[width=0.49\textwidth]{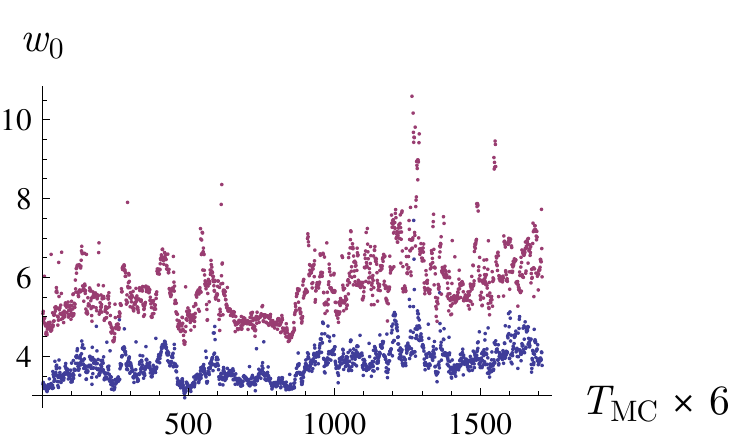}}
\subfigure[$w_0^{0.1}$ versus $w_0^{0.4}$]
{\includegraphics[width=0.49\textwidth]{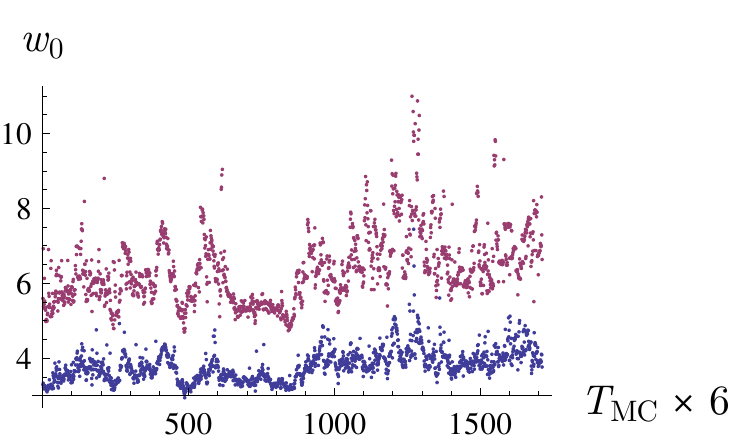}}
\caption{Comparison of the Monte Carlo history (a) of $w_0^{0.1}$ (blue)
with $w_0^{0.3}$ (red) and (b) of $w_0^{0.1}$ (blue) with $w_0^{0.4}$ (red),
on a $32^3 \times 64$ lattice, $\beta=1.9$ and $\kappa=0.14435$. The
magnitude of the peaks increases drastically when the reference value to set
the $w_0$ scale is larger.}
\label{comparison_mc_history_w0}
\end{figure}

Wilson flow scales depend implicitly on the reference value chosen in
Eq.~\ref{definition_w0}. Small reference values will potentially produce
large lattice artefacts on the final results, while $w_0$ and $t_0$ will be
affected by non-negligible finite volume effects for larger reference
values. Let us define the scale $w_0^u$ to be the square root of the flow
time when the condition
\begin{equation}\label{generalized_w0}
 t \frac{d}{dt} t^2 \langle E(t) \rangle = u
\end{equation}
is satisfied. By varying $u$ one can study how the autocorrelations are
affected by different choices of the reference value. Here and in the
following we set $w_0 \equiv w_0^{0.3}$. In
Fig.~\ref{comparison_mc_history_w0} the Monte Carlo histories of
$w_0^{0.1}$, $w_0^{0.3}$ and $w_0^{0.4}$ are compared. When the value of $u$
is small the fluctuations and spikes are significantly reduced. On the other
hand, when the value of $u$ is increased the spikes become even more pronounced.

\begin{figure}[!b]
\centering
\subfigure[Topological distribution of $w_0^{0.3}$]
{\includegraphics[width=0.49\textwidth]{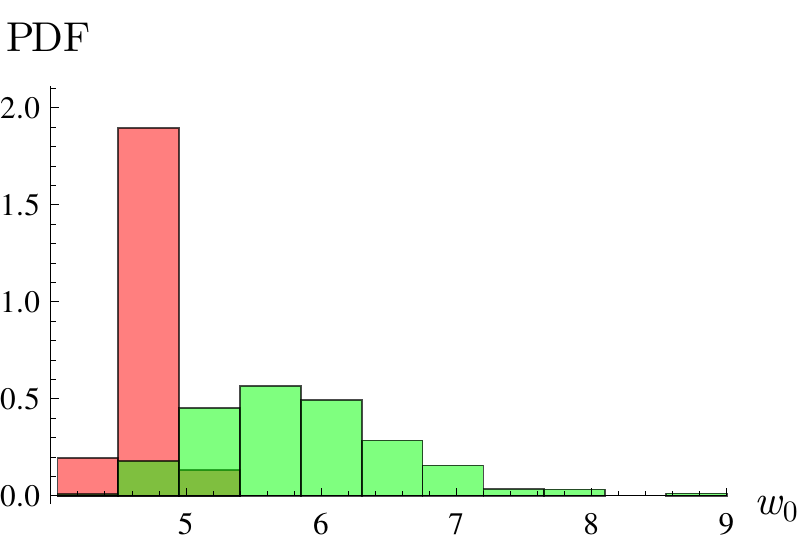}}
\subfigure[Topological distribution of $w_0^{0.4}$]
{\includegraphics[width=0.49\textwidth]{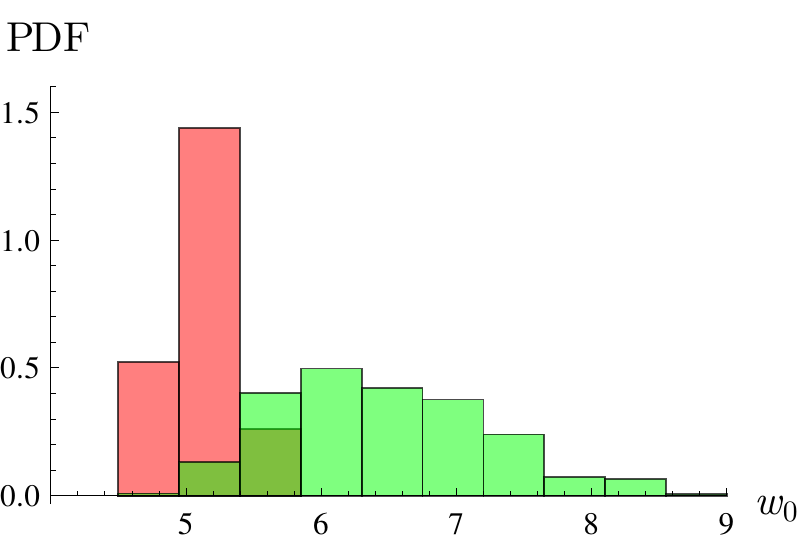}}
\caption{Probability distribution function of (a) $w_0^{0.3}$ and of (b)
$w_0^{0.4}$ restricted to the topological sector $|Q_\textrm{top}| = 1$
(green) and $|Q_\textrm{top}| = 4$ (red) on a $32^3 \times 64$ lattice,
$\beta=1.9$ and $\kappa=0.14435$.}
\label{w0_topological_sector_distributions}
\end{figure}
Increasing the value of $u$ leads to a larger flow time needed to match the
condition (\ref{generalized_w0}), which means that a stronger smoothing
induced by the flow equation is applied to the configurations. Large flow
times will stronger remove ultraviolet fluctuations, and the system will be
brought towards a classical configuration, as observed in
section \ref{topological_charge_sec}. Therefore one might argue that spikes and large fluctuations are
related to topological effects. Using the results presented in
section \ref{topological_charge_sec} we have been able to compute the value of $w_0^u$ restricted only to
configurations with a fixed definite topological charge.
\begin{figure}[tbh]
\centering
\includegraphics[width=0.59\textwidth]{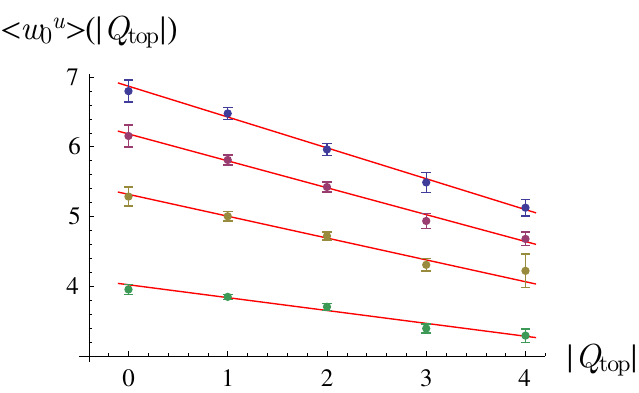}
\caption{Linear fit of the dependence of $w_0^{u}$ on the topological charge
for $u$ equal to 0.1 (green), 0.2 (yellow), 0.3 (purple) and 0.4 (blue) for
a lattice $32^3 \times 64$, $\beta=1.9$ and $\kappa=0.14435$.}
\label{w0_topological_sector_fit_1900b}
\end{figure}

The distributions of $w_0^{0.4}$ and $w_0^{0.3}$ are shown for the same run
in Fig.~\ref{w0_topological_sector_distributions} for two selected
topological sectors, $|Q_\textrm{top}| = 1$ and $|Q_\textrm{top}| = 4$. The
distribution of $w_0^{0.3}$ restricted to the topological sector
$|Q_\textrm{top}| = 1$ is rather broad and the average value is larger
than for the distribution in the topological sector $|Q_\textrm{top}| = 4$. The
same behaviour appears for the restricted distributions of $w_0^{0.4}$, but
with a slightly larger difference between the two mean values of the
distributions. This result clearly shows that there is a correlation
between the value of $w_0$ and the topological charge, not only in terms
of its mean value but also in terms of its distribution.
The largest fluctuations observed in Fig.~\ref{mc_history_w0_03} are
produced by the configurations at low values of the topological charge,
where the distribution of $w_0$ is broad. The long
periodicity is induced by the transitions during the Monte Carlo update time
between topological sectors around zero, characterised by large expectation
value of $w_0$, and topological sectors far from the origin with a small
mean value of $w_0$. The Monte Carlo history restricted to a given topological 
sector is presented in Fig.~\ref{plot_history_restricted} in the Appendix.

\begin{figure}[t]
\centering
\subfigure[$s$ versus $u$]
{\includegraphics[width=0.45\textwidth]{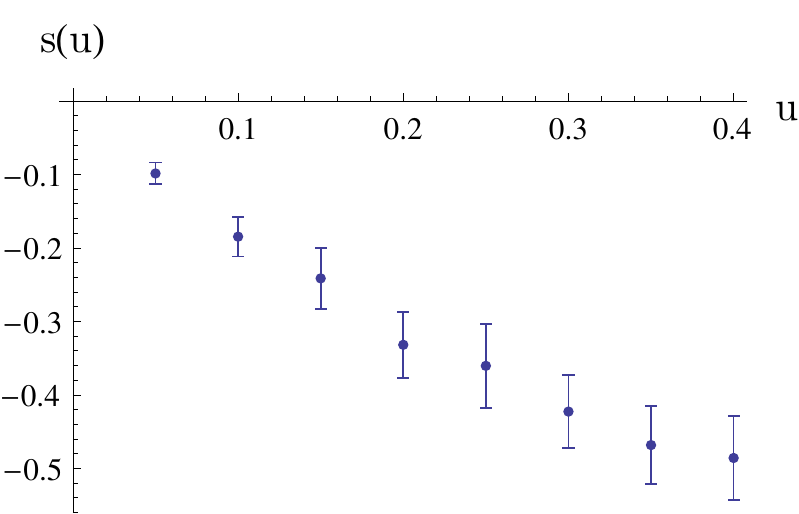}
\label{w0_m_u_dependence}}
\subfigure[$s$ versus $(w_0 m_\pi)^2$]
{\includegraphics[width=0.49\textwidth]{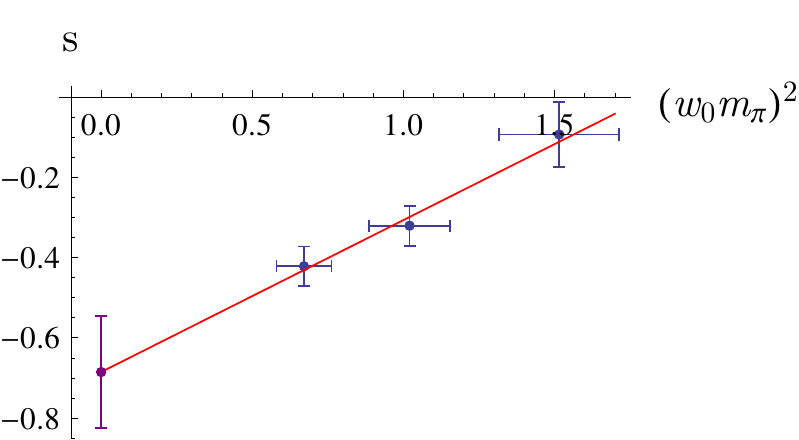}
\label{topological_slope_1900b}}
\caption{(a) Slope coefficient $s$ as a function of the reference value $u$
for the lattice $32^3\times64$, $\beta=1.9$ and $\kappa=0.14435$. (b) Slope
coefficient $s$ as a function of $(w_0 m_\pi)^2$ at $\beta=1.9$ for
$w_0^{0.3}$. The value of $s$ linearly extrapolated to the chiral limit is
$s((w_0 m_\pi)^2 = 0) = -0.69(14)$.}
\end{figure}

In Fig.~\ref{w0_topological_sector_fit_1900b} we present the expectation
value of $w_0^u$ restricted to the various topological sectors for four
different values of $u$, in the same run on a $32^3 \times 64$ lattice, with
$\beta=1.9$ and $\kappa=0.14435$. The behaviour of $w_0^u(|Q_\textrm{top}|)$
is approximately linear for all $u$ but it has a steeper slope when the
reference scale $u$ is larger. We have used a linear fit of the form
\begin{equation}
 \langle w_0^u \rangle (|Q_\textrm{top}|) = s |Q_\textrm{top}| + q.
\end{equation}
The resulting slope coefficients $s$ are presented as a function of $u$ in
Fig.~\ref{w0_m_u_dependence}. The modulus of the slope $s$ increases 
increasing $u$. This means that the dependence of $w_0^u$ on the topology is
stronger when $u$ is larger. This behaviour confirms our previous claim about
the topological origin of the spikes in Fig.~\ref{comparison_mc_history_w0}:
when $u$ is large the smoothing effects of the Wilson flow are large and the
configuration is driven towards a classical one where the influence of the
topology is stronger. As a result, the integrated autocorrelation time of $w_0^{0.4}$ is
around 800 $T_{MC}$, approximately three times larger than the
autocorrelation time of $w_0^{0.1}$, which is around 300 $T_{MC}$. We
have also investigated the dependence of $s$ on the adjoint pion
mass squared, observing that it increases as one approaches the chiral
limit, see Fig.~\ref{topological_slope_1900b}.

\begin{figure}[t]
\centering
\subfigure[$w_0^{u}(|Q_\textrm{top}|)$, lattice $32^3 \times 64$]
{\includegraphics[width=0.48\textwidth]{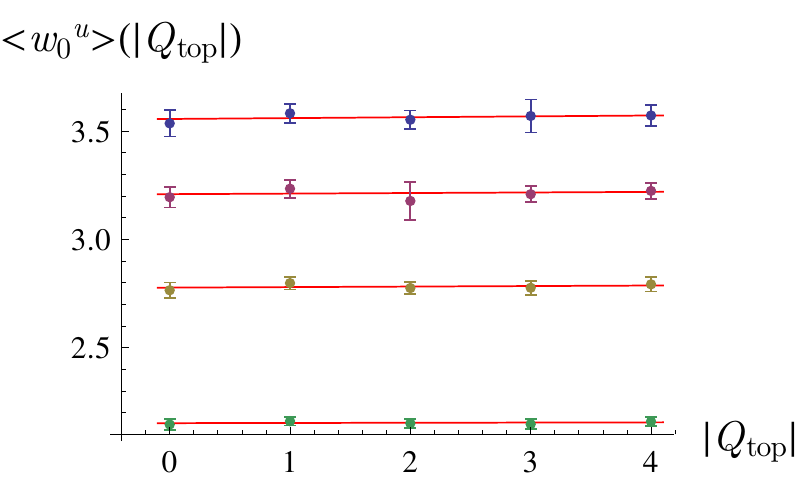}
\label{w0_topological_sector_fit_1750b}}
\subfigure[$w_0^{u}(|Q_\textrm{top}|)$, lattice $16^3 \times 36$]
{\includegraphics[width=0.48\textwidth]{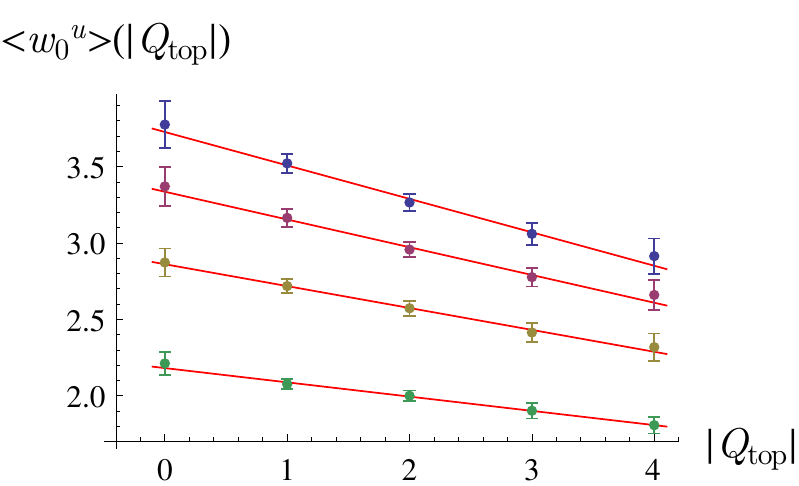}
\label{w0_topological_sector_fit_1750b_reduced_volume}}
\caption{The same as Fig.~\ref{w0_topological_sector_fit_1900b}, but for (a)
the lattice $32^3\times 64$, $\beta=1.75$ and $\kappa=0.1494$; (b) the
lattice $16^3\times 36$, $\beta=1.75$ and $\kappa=0.1490$.}
\end{figure}

The dependence of the flow scale on the topological charge can be
interpreted as a finite volume effect \cite{Brower:2003yx,Aoki:2007ka}. To
address this point we repeated the same systematic analysis on the lattice
$32^3 \times 64$ at $\beta=1.75$, where the physical volume is approximately
seven times larger than at $\beta=1.9$. The dependence of the various
$w_0^u$ on $|Q_{\textrm{top}}|$ is presented in
Fig.~\ref{w0_topological_sector_fit_1750b}. As the figure shows, in this
large physical volume the dependence of the flow scales on the topology
completely disappears. If instead, the physical volume is shrunk again by
simulating on a $16^3 \times 36$ lattice at the same $\beta=1.75$, the
observables $w_0^u$ appear to depend on the topological charge
$|Q_{\textrm{top}}|$ as before, see
Fig.~\ref{w0_topological_sector_fit_1750b_reduced_volume}. 
Note that the value of $\kappa$ used in the smaller volume is smaller
than the first case, but according to Fig.~\ref{topological_slope_1900b}
this should even reduce the slope. This demonstrates
that finite volume effects are the origin of the dependence of $w_0$ on the
topological charge in the runs on the finer lattices at $\beta=1.9$.

\section{Conclusions}

We have presented a detailed analysis of the Wilson flow observables $w_0$, 
used to set the scale alternatively to the Sommer parameter $r_0$. 
The same analysis has been done also for $t_0$ reaching similar conclusions. 
In finite volumes we observed a substantial dependence of $w_0$ on
the topological charge, in agreement with the previous discussion on this
topic for the Sommer parameter $r_0$ of
Ref.~\cite{Bruckmann:2009cv,Aoki:2008tq}.

Scales based on the Wilson flow require a delicate fine-tuning to correctly
handle finite volume effects and errors due to lattice artefacts. A result
free of \emph{topological} finite volume effects can be ensured if there is no coupling between the scale and
the topological charge, up to the statistical precision. 
 The final result has fairly small statistical and systematic
errors, therefore $w_0$ can be used to set the scale in extrapolations to the
chiral and to the continuum limit. Our observations support the use of small
flow times to set the scale, at least for our model and within our present
precision: the ratio of $w_0^u$ and $w_0^{0.3}$ is flat for $u \gtrsim 0.1$
(see Tab.~\ref{tab:wuresults} and Fig.~\ref{scaling_w0u} in the Appendix).

The results that have been presented may be different for other theories, in
particular they may depend on the number of colours $N_c$, on the number of
fermions $N_f$ and on the fermion representation. We believe, however, that
a dependence of the scale on topology emerges for sufficient large reference
scales independently of the theory. We therefore encourage systematic
studies in this direction, in particular 
considering that there are proposals, like in~\cite{Sommer:2014mea}, to 
increase the value of the reference flow time.

\section{Acknowledgments}

The authors gratefully acknowledge the Gauss Centre for Supercomputing (GCS) for providing computing time for a GCS Large Scale Project on the GCS share of the supercomputer JUQUEEN at J\"ulich Supercomputing Centre (JSC). GCS is the alliance of the three national supercomputing centres HLRS (Universit\"at Stuttgart), JSC (Forschungszentrum J\"ulich), and LRZ (Bayerische Akademie der Wissenschaften), funded by the German Federal Ministry of Education and Research (BMBF) and the German State Ministries for Research of Baden-W\"urttemberg (MWK), Bayern (StMWFK) and Nordrhein-Westfalen (MIWF).


\newpage
\appendix

\section{Tables and additional figures}

\begin{table}[h]
  \centering
  \begin{tabular}{cccccccc}
  \hline
  Volume & $\beta$ & $\kappa$ & $a m_{\pi}$ & $\sqrt{t_0}/a$ & $w_0/a$ & $\tau({t_0})$ \\
  \hline
  $24^3 \times 48$ & 1.60 & 0.15500 & 0.5788(16) & 1.5672(13) & 1.5102(14) & 21 \\ 
  $24^3 \times 48$ & 1.60 & 0.15700 & 0.3264(23) & 1.7904(11) & 1.7292(37) & 10 \\ 
  $24^3 \times 48$ & 1.60 & 0.15750 & 0.2015(93) & 1.8986(53) & 1.8410(63) & 42 \\ 
  \hline
  $32^3 \times 64$ & 1.75 & 0.14900 & 0.2385(4)  & 3.1438(67) & 2.9838(59) & 50 \\
  $32^3 \times 64$ & 1.75 & 0.14920 & 0.2035(5)  & 3.270(17)  & 3.097(25)  & 45\\
  $32^3 \times 64$ & 1.75 & 0.14940 & 0.1604(15) & 3.362(15)  & 3.205(20)  & 35 \\
  $32^3 \times 64$ & 1.75 & 0.14950 & 0.1294(24) & 3.551(36)  & 3.413(40)  & 65\\
  \hline
  $32^3 \times 64$ & 1.90 & 0.14387 & 0.2123(4) & 5.73(13) & 5.57(19) & 440 \\
  $32^3 \times 64$ & 1.90 & 0.14415 & 0.1742(4) & 5.71(12) & 5.49(11) & 296 \\
  $32^3 \times 64$ & 1.90 & 0.14435 & 0.1413(6) & 5.96(12) & 5.76(14) & 502 \\
  \hline
  \end{tabular}
  \caption{Results for the adjoint pion mass $m_{\pi}$, the scales $t_0$,
  $w_0$ and the autocorrelation time $\tau({t_0})$ of $t_0$.}
  \label{tab:w0results}
\end{table}

\begin{table}[h]
  \centering
  \begin{tabular}{ccccc}
  \hline
  Run & Volume & $\beta$ & $\kappa$ & $(a^4 \chi_{\textrm{top}})  \times 10^{-6}$ \\
  \hline
  A1 & $24^3 \times 48$ & 1.60 & 0.15500 & 160(19) \\ 
  A2 & $24^3 \times 48$ & 1.60 & 0.15700 & 102(9) \\ 
  A3 & $24^3 \times 48$ & 1.60 & 0.15750 & 85(7) \\ 
  \hline
  B1 & $32^3 \times 64$ & 1.75 & 0.14900 & 17(2) \\
  B2 & $32^3 \times 64$ & 1.75 & 0.14920 & 12(1) \\
  B3 & $32^3 \times 64$ & 1.75 & 0.14940 & 11(1) \\
  B4 & $32^3 \times 64$ & 1.75 & 0.14950 & 10(3) \\
  \hline
  C1 & $32^3 \times 64$ & 1.90 & 0.14387 & 1.01(14) \\
  C2 & $32^3 \times 64$ & 1.90 & 0.14415 & 1.59(18) \\
  C3 & $32^3 \times 64$ & 1.90 & 0.14435 & 1.02(8) \\
  \hline
  \end{tabular}
  \caption{Results for the topological susceptibility.}
  \label{tab:chiresults}
\end{table}

\begin{landscape}
\begin{table}[h]
  \centering
  \begin{tabular}{ccccccccc}
  \hline
  Run & $w_0^{0.05}$ & $w_0^{0.10}$ & $w_0^{0.15}$ & $w_0^{0.20}$ & $w_0^{0.25}$ & $w_0^{0.30}$ & $w_0^{0.35}$ & $w_0^{0.40}$ \\
  \hline
  A1 & 0.6419(10) & 0.9587(15) & 1.1420(17) & 1.2844(19) & 1.4048(22) & 1.5104(21) & 1.6046(26) & 1.6901(26) \\ 
  A2 & 0.7767(15) & 1.1136(22) & 1.3192(27) & 1.4787(34) & 1.6122(40) & 1.7290(43) & 1.8319(46) & 1.9268(44) \\ 
  A3 & 0.8341(23) & 1.1853(35) & 1.4048(41) & 1.5760(47) & 1.7182(53) & 1.8419(60) & 1.9511(58) & 2.0495(68) \\ 
  \hline
  B1 & 1.5087(42) & 2.0058(51) & 2.3331(56) & 2.5878(60) & 2.8011(62) & 2.9838(59) & 3.1495(62) & 3.3012(62) \\
  B2 & 1.5610(84) & 2.082(15) & 2.417(19) & 2.684(19) & 2.904(23) & 3.097(25) & 3.256(27) & 3.424(25) \\
  B3 & 1.6113(74) & 2.151(12) & 2.504(13) & 2.779(17) & 3.005(18) & 3.205(20) & 3.385(22) & 3.551(25) \\
  B4 & 1.6957(15) & 2.269(24) & 2.651(28) & 2.955(34) & 3.205(37) & 3.413(41) & 3.623(43) & 3.793(42) \\
  \hline
  C1 & 2.734(46) & 3.687(81) & 4.35(10) & 4.82(14) & 5.24(16) & 5.60(18) & 5.90(18) & 6.19(18)\\
  C2 & 2.717(40) & 3.666(85) & 4.31(11) & 4.76(13) & 5.13(13) & 5.49(13) & 5.78(15) & 6.05(14) \\
  C3 & 2.855(44) & 3.822(81) & 4.47(11) & 4.97(11) & 5.40(14) & 5.76(14) & 6.08(15) & 6.38(15) \\
  \hline
  \end{tabular}
  \caption{Results for the scale $w_0^u$.}
  \label{tab:wuresults}
\end{table}
\end{landscape}
\newpage
\begin{figure}[t]
\centering
\includegraphics[width=0.45\textwidth]{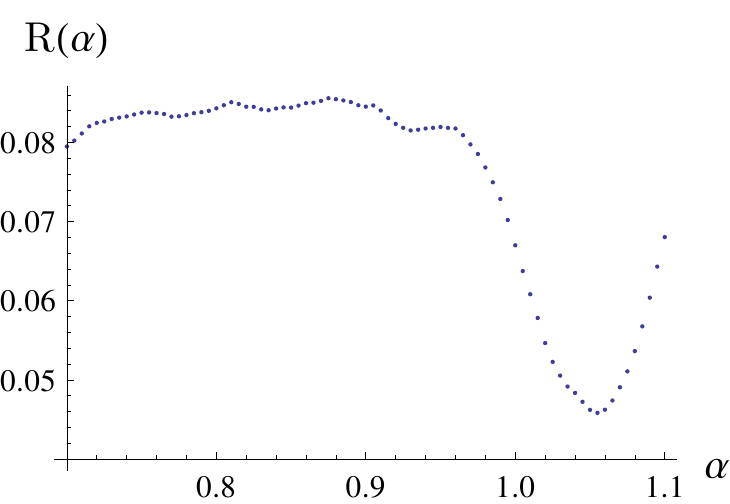}
\caption{Plot of $R(\alpha)$ for the lattice $32^3 \times 64$ at
$\beta=1.75$ and $\kappa=0.1494$. The minimum of $R(\alpha)$ is located at
$\alpha=1.055$.}\label{ralpha}
\end{figure}

\begin{figure}[tbh]
\centering
\subfigure[Topological sector 0]
{\includegraphics[width=0.49\textwidth]{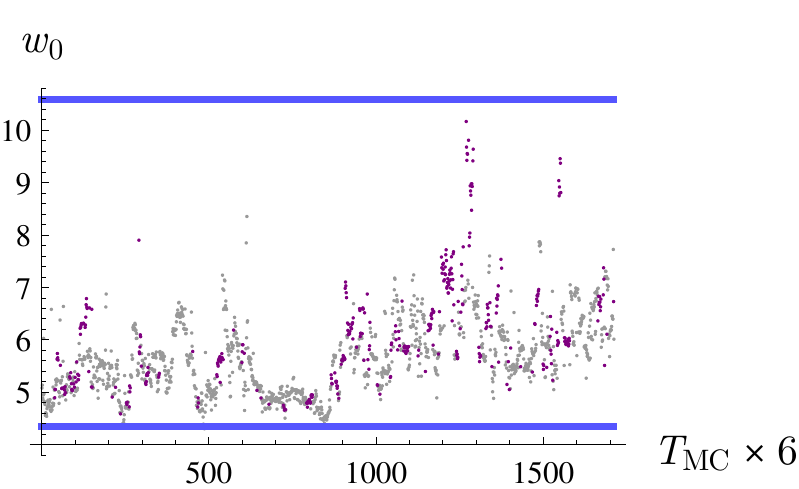}}
\subfigure[Topological sector 1]
{\includegraphics[width=0.49\textwidth]{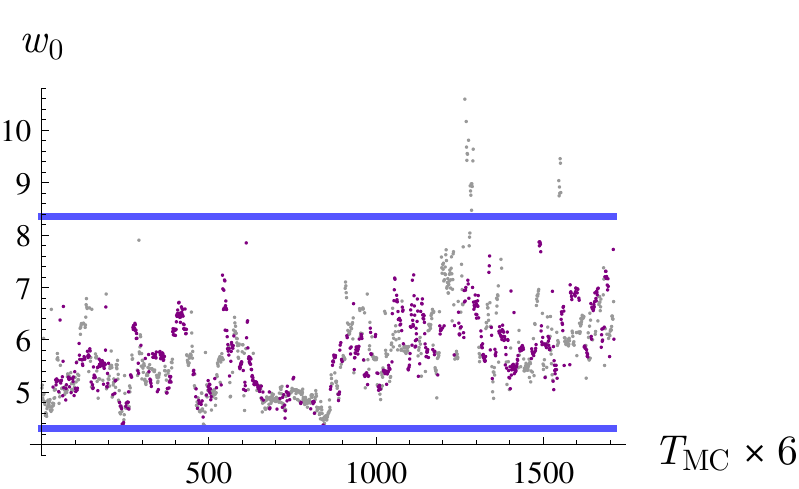}}
\subfigure[Topological sector 2]
{\includegraphics[width=0.49\textwidth]{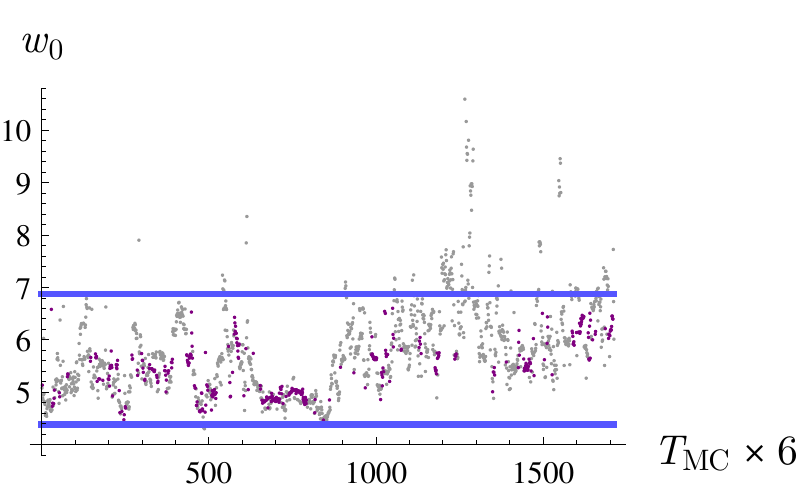}}
\subfigure[Topological sector 3]
{\includegraphics[width=0.49\textwidth]{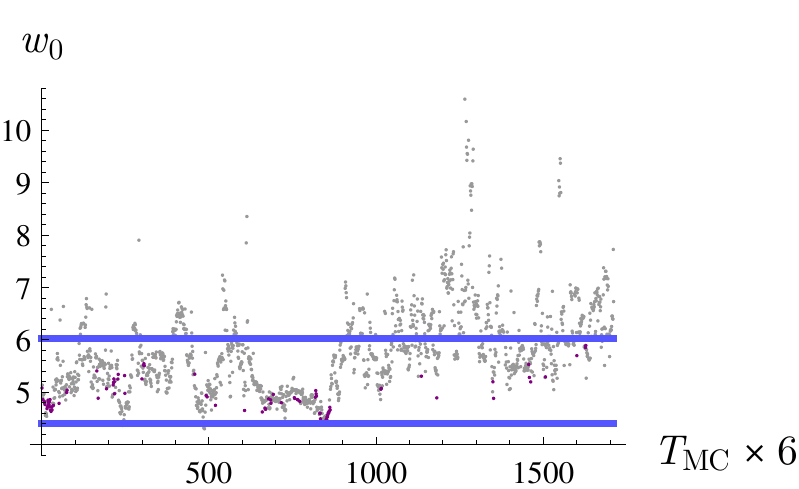}}
\caption{Monte Carlo history of the $w_0$ on a $32^3 \times 64$ lattice,
$\beta=1.9$ and $\kappa=0.14435$. The purple points highlight the value of $w_0$ only for configurations characterised by a given topological sector; the blue lines mark its maximal and minimal value.}\label{plot_history_restricted}
\end{figure}

\begin{figure}[tbh]
\centering
\subfigure[$w_0^u/w_0^{0.3}$ for $u=0.05$ (blue), $u=0.15$ (purple), $u=0.25$ (yellow), $u=0.35$ (green).]
{\includegraphics[width=0.85\textwidth]{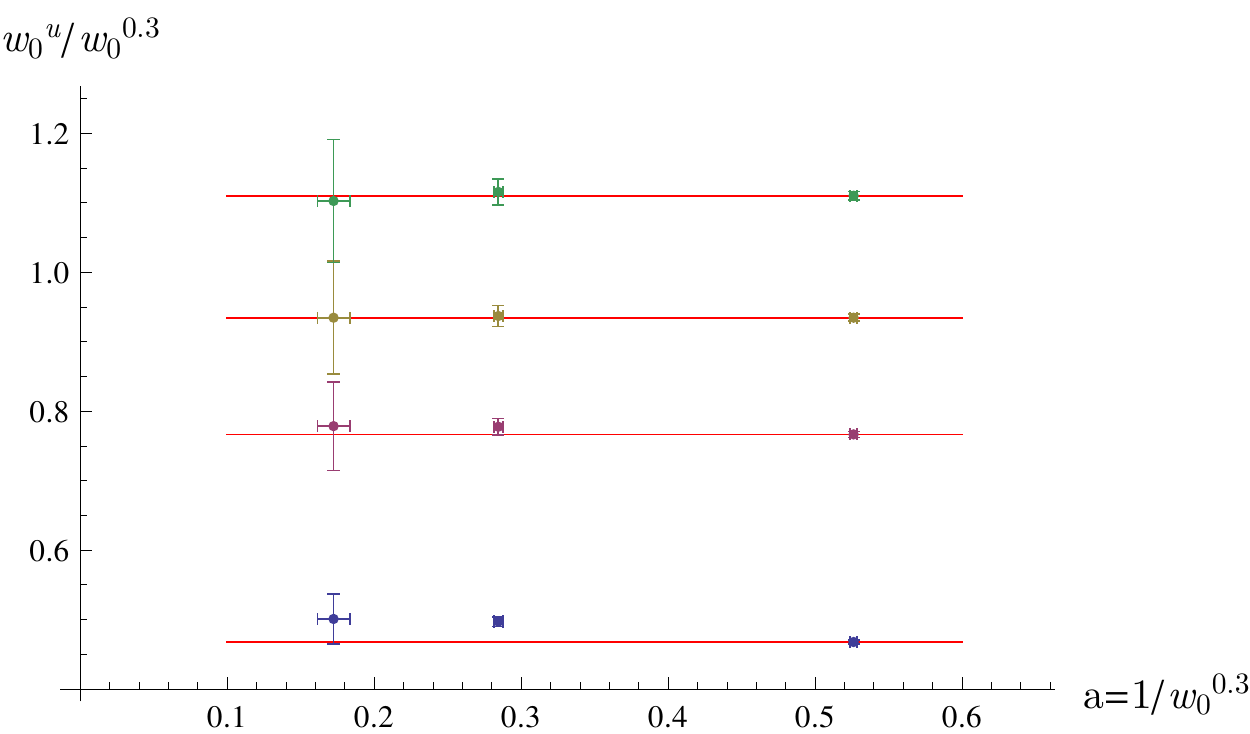}}
\subfigure[$w_0^u/w_0^{0.3}$ for $u=0.10$ (blue), $u=0.20$ (purple), $u=0.40$ (yellow).]
{\includegraphics[width=0.85\textwidth]{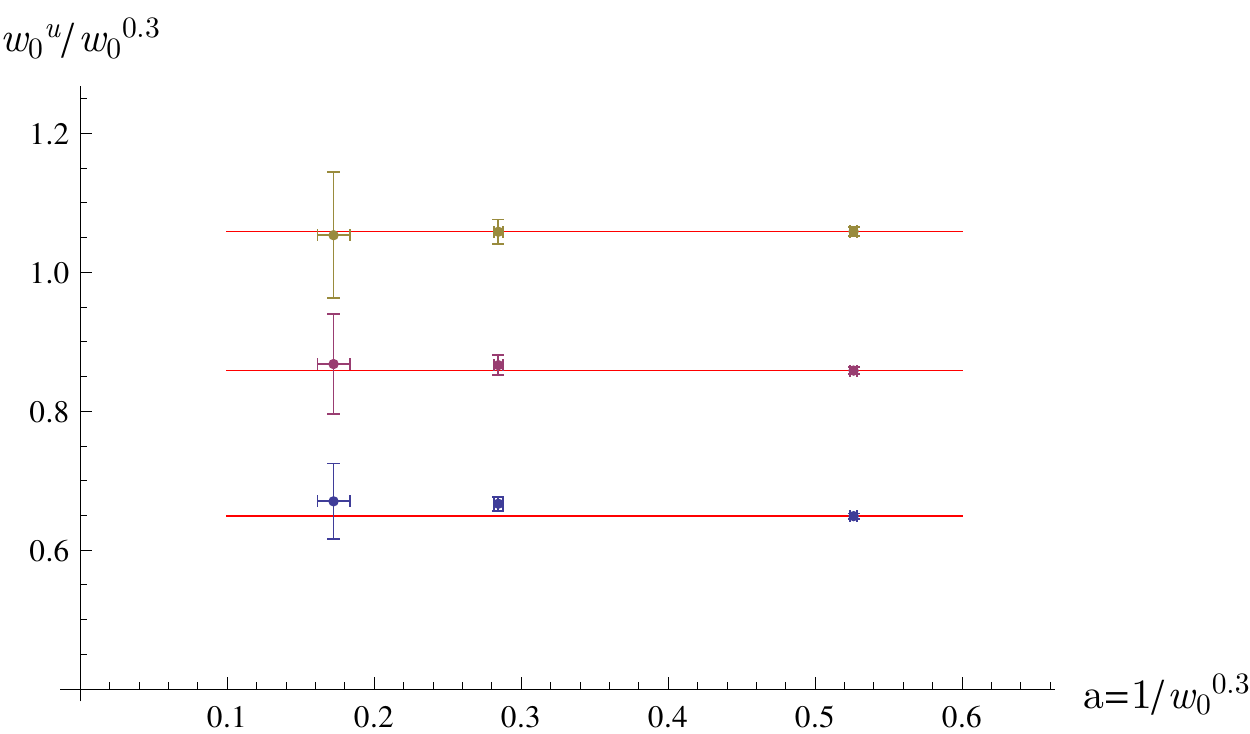}}
\caption{Scaling of the various $w_0^u$ with respect to $w_0^{0.3}$ as a function of the lattice spacing. The scaling is flat within the errors for $u \gtrsim 0.15$.}\label{scaling_w0u}
\end{figure}

\end{document}